\documentclass[runningheads,a4paper]{llncs}

\usepackage{amssymb}
\usepackage{amsmath}
\usepackage{graphics, graphicx}
\usepackage[]{appendix}
\graphicspath{ {./images/} }

\usepackage[table,pdftex]{xcolor}
\usepackage{xspace}
\usepackage[colorlinks=true,citecolor=blue,urlcolor=black]{hyperref}
\usepackage{subfig}
\usepackage[colorinlistoftodos]{todonotes}
\usepackage{multirow, tabularx}
\renewcommand{\arraystretch}{1.10}
\newcolumntype{C}{@{\extracolsep{0.2cm}}c@{\extracolsep{0pt}}}%
\usepackage{booktabs}       

\usepackage{pifont}
\hypersetup{
    colorlinks,
    linkcolor={blue},
    citecolor={blue},
    urlcolor={blue}
}

\begin{document}

\mainmatter  
\title{Image-level Harmonization of Multi-Site Data using Image-and-Spatial Transformer Networks}
\titlerunning{Image-level Harmonization of Multi-Site Data using ISTNs}

\author{R.~Robinson$^1$, Q.~Dou$^1$, D.~C.~Castro$^1$, K. Kamnitsas$^1$, M.~de~Groot$^2$, R.M.~Summers$^3$, D.~Rueckert$^1$, B.~Glocker$^1$}

\authorrunning{Robinson et al.}


\institute{
$^1$ BioMedIA, Department of Computing, Imperial College London, UK\\
$^2$ Research \& Development, GlaxoSmithKline, UK\\
$^3$ Department of Radiology and Imaging Sciences, Clinical Center,\\National Institutes of Health, USA\\%
}

\maketitle

\begin{abstract} We investigate the use of image-and-spatial transformer networks (ISTNs) to tackle domain shift in multi-site medical imaging data. Commonly, domain adaptation (DA) is performed with little regard for explainability of the inter-domain transformation and is often conducted at the feature-level in the latent space. We employ ISTNs for DA at the image-level which constrains transformations to explainable appearance and shape changes. As proof-of-concept we demonstrate that ISTNs can be trained adversarially on a classification problem with simulated 2D data. For real-data validation, we construct two 3D brain MRI datasets from the Cam-CAN and UK Biobank studies to investigate domain shift due to acquisition and population differences. We show that age regression and sex classification models trained on ISTN output improve generalization when training on data from one and testing on the other site.
\end{abstract}


\section{Introduction}

Domain shift (DS) concerns the problem of mismatch between the statistics of the training data used for model development and the statistics of the test data seen after model deployment. DS can cause significant drops in predictive performance, which has been observed in almost all recent imaging challenges when final test data was coming from different clinical sites \cite{crimi2019a}. DS is a major hurdle for successfully translating predictive models into clinical routine.

Acquisition and population shift are two common forms of DS that appear in medical image analysis \cite{castro2019causality}. Acquisition shift is observed due to differences in imaging protocols, modalities or scanners. Such a shift will be observed even if the same subjects are scanned. Population shift occurs when cohorts of subjects under investigation exhibit different statistics, e.g., varying demographics or disease prevalence. It is not uncommon for both types of DS to occur simultaneously, in particular in multi-center studies. It is essential to tackle DS in machine learning to perform reliable analysis of large populations across sites and to avoid introducing biases into results. Recent work has shown that even after careful pre-processing, site-specific differences remain in the images \cite{wachinger2019a,glocker2019multisite}. While methods like ComBat \cite{combat2018} aim to harmonize image-derived measurements, we focus on the images themselves.

One solution is domain adaptation (DA), a transductive \cite{pan2010a} transfer learning technique that aims to modify the source domain's marginal distribution of the feature space such that it resembles the target domain. In medical imaging, labelled data is scarce and typically unavailable for the target domain. It is also unlikely to have the same subjects in both domains. Thus, we focus on `unsupervised' and `unpaired' DA, wherein labelled data is available only in the source domain and no matching samples exist between source and target.

Many DA approaches focus on learning domain-invariant feature representations, by either forcing latent representations of the inputs to follow similar distributions, or `disentangling' domain-specific features from generic features \cite{yang2019a}. This can be achieved with some divergence measure based on data statistics or by training adversarial networks to model the divergence between the feature representations \cite{Yan2019a}. These methods have been applied to brain lesions \cite{kamnitsas2017unsupervised} and tumours \cite{dai2019a} in MRI, and in contrast to non-contrast CT segmentation \cite{Sandfort2019}

While these approaches seem appealing and have shown some success, they lack a notion of explainability as it is difficult to know what transformations are applied to the feature space. Additionally, although the learned task model may perform equally well on both domains, it is not guaranteed to perform as well as separate models trained on the individual domains.

We explore model-agnostic DA by working at the image level. Our approach is based on domain mapping (DM), which aims to learn the pixel-level transformations between two image domains, and includes techniques such as style transfer. Pix2Pix \cite{isola2017a} (supervised) and CycleGAN \cite{zhu2017a} (unsupervised) take images from one domain through some encoder-decoder architecture to produce images in the new domain. The method in \cite{Yan2019a} uses CycleGAN to improve segmentation across scanners and applies DA at both image and feature levels, thus losing interpretability. It does not decompose the image and spatial transformations.

Methods for DM primarily use UNet-like architectures to learn image-to-image transformations that are easier to interpret, as one can visually inspect the output. For medical images of the same anatomy, but from different scanners, we assume that domain shift manifests primarily in appearance changes (contrast, signal-to-noise, resolution) and anatomical variation (shape changes), plus further subtle variations caused by image reconstruction or interpolation.\\

\noindent\textbf{Contributions:} We propose the use of image-and-spatial transformer networks (ISTNs) \cite{ISTN2019} to tackle domain shift at image-feature level in multi-site imaging data. ISTNs separate and compose the transformations for adapting appearance and shape differences between domains. We believe our approach is the first to use such an approach with retraining of the downstream task model on images transferred from source to target. We show that ISTNs can be trained adversarially in a task model-agnostic way. The transferred images can be visually inspected, and thus, our approach adds explainability to domain adaptation---which is important for validating the plausibility of the learned transformations. Our results demonstrate the successful recovery of performance on classification and regression tasks when using ISTNs to tackle domain shift. We explore both unidirectional and bidirectional training schemes and compare retraining the task model from scratch versus finetuning. We present proof-of-concept results on synthetic images generated with Morpho-MNIST \cite{castro2019morphomnist} for a 3-class classification task. Our method is then validated on real multi-site data with 3D T1-weighted brain MRI. Our results indicate that ISTNs improve generalization and predictive performance can be recovered close to single-site accuracy.
\begin{figure}[tb!]
	\includegraphics[width=\linewidth]{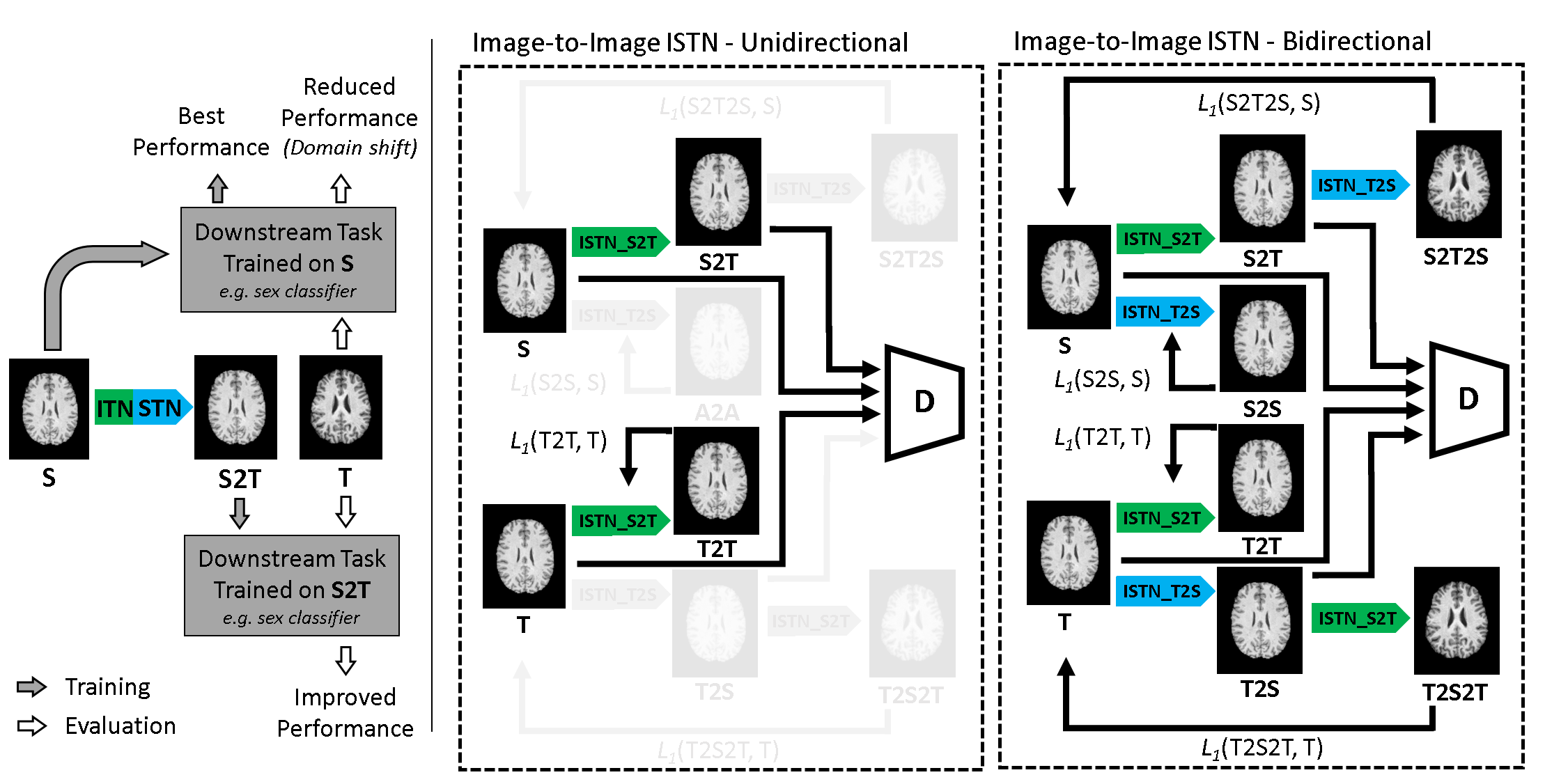}
	\caption{\footnotesize (left) The domain shift problem can be mitigated by retraining or finetuning a task model on images $S2T$. (Middle) The ISTN is trained adversarially such that the discriminator $D$ becomes better at identifying real ($S$ and $T$) and transformed ($S2T$) images. The ISTN simultaneously produces better transformations $S2T$ of $S$ that look more like the images $T$. The training process can also be done bidirectionally (right).}
		\label{fig:fig1}
\end{figure}

\section{Method}
\label{sec:method}

We propose adversarial training of ISTNs to perform model-agnostic DA via explicit appearance and shape transformations between the domains. We explore unidirectional and bidirectional training schemes as illustrated in Figure~\ref{fig:fig1}.
\\
\\
\noindent\textbf{Models.} ISTNs have two components: an image transformer network (ITN) and a spatial transformer network (STN) \cite{jaderberg2015spatial,ISTN2019}. Here, we additionally require a discriminator model for adversarial training of the ISTN.

\textit{ITN:} The ITN performs appearance transformations such as contrast and brightness changes, and other localised adaptations at the image-level. A common image-to-image (I2I) translation network based on UNet with residual skip connections can be employed. We use upsample-convolutions to reduce chequerboard artifacts compared with transposed convolution. We use batch normalization, dropout layers and ReLU activations with a final $\tanh$ activation for the output. All input images are pre-normalized to the $[-1,1]$ intensity range.

\textit{STN:} We experiment with both the affine and B-spline STNs described in the original ISTN paper. Affine STNs learn to regress the parameters of linear spatial transforms with translation, rotation, scaling, and shearing. B-spline STNs regress control point displacements. Linear interpolation is used throughout. Note that in this work, Affine and B-Spline STNs are considered independently and are not composed.

\textit{Discriminator:} In both Morpho-MNIST and brain MRI experiments, we use a standard fully-convolutional classification network with instance normalization, dropout layers and a sigmoid output.

\textit{Task models:} The employed classifiers and regressors follow the same fully-convolutional structure as the discriminator, reducing the dimensions of the input images to a multi-class or continuous value prediction, depending on the task. We use cross-entropy or mean-squared error loss functions, respectively.

Appendices \ref{appendix:appendixC} and \ref{appendix:appendixD} provide details about the architectures of different networks. All implementations are in PyTorch \cite{NEURIPS2019_9015} with code available online.\footnote{\url{https://github.com/mlnotebook/domain_adapation_istn}}.
\\
\\
\noindent\textbf{Training.} The output from the ITN is directly fed into the STN. They are then composed into a single ISTN unit, and are trained jointly end-to-end.
\textit{Discriminator}: The images $S$ (from the source domain) are passed through the ISTN to generate images $S2T$, where $T$ indicates images from the target domain. Next, the $S2T$ are passed through the discriminator $D_T$ to yield a score in the range $(0,1)$ denoting whether the image is a real sample from domain $T$ or a transformed one. The discriminator is trained by minimizing the binary cross-entropy loss $\mathcal{L}_{bce}$ between the predicted and true domain labels. Eq.~(\ref{eq:dis_loss_1}) shows the total discriminator loss. Soft labels for the true domain are used to stabilize early training of the discriminator. We replace the hard `0' and `1' domain labels by random uniform values in the ranges $[0.00,0.03]$ and $[0.97,1.00]$, respectively.

\textit{ISTN}: The ISTN is trained as a generator. The ISTN output $S2T$ is passed through the discriminator and forced to be closer to domain $T$ by computing the adversarial loss $\mathcal{L}_{adv} = \mathcal{L}_{bce}(D_T(S2T), 1)$. Soft labels are also used here. We expect that when images $T$ are passed through the ISTN, the output $T2T$ should be unchanged as it is already in domain $T$. This is enforced by the identity loss $\mathcal{L}_{idt} = \ell_{1}(T, T2T)$ acting on image intensities of $T$ and $T2T$. A weighting factor $\lambda$ is applied to $L_{idt}$ giving the total loss function for the ISTN in Eq.~(\ref{eq:istn_cycleloss_a2b})c.

We compare with the CycleGAN \cite{CycleGAN2017} training approach, which trains both directions simultaneously using two ISTNs ($\operatorname{ISTN}_{S2T}$ and $\operatorname{ISTN}_{T2S}$) and two discriminators ($D_S$ and $D_T$). The CycleGAN introduces the cycle-consistency term to $\mathcal{L}_{istn}$ such that when $\operatorname{ISTN}_{T2S}$ is used to transform $S2T$, the result $S2T2S$ is forced to be close to $S$. Figure~\ref{fig:fig1} shows the two ISTNs, their outputs and associated losses. The loss functions for $\operatorname{ISTN}_{S2T}$ are shown in Eq.~(\ref{eq:istn_cycleloss_a2b}). Optimization is done using the Adam optimizer.

\noindent\textbf{Downstream Tasks:} The goal of our work is to demonstrate that such explicit appearance and spatial transformations via ISTNs can successfully tackle DS in certain applications. Ideally, we would like to observe that the performance of a predictor trained on $S2T$ and tested on $T$ can recover to single-site performance. To demonstrate this, prior to training the ISTN, we train a task model (\textit{e.g.} classifier or regressor) $\mathcal{T}_S$ on domain $S$. The performance of $\mathcal{T}_S(S)$ is likely to be our `best performance' whilst $\mathcal{T}_S(T)$ will degrade due to DS. During ISTN training, we simultaneously re-train $\mathcal{T}_S$ on the ISTN output of $S2T$. This model $\mathcal{T}_{S2T}$ is trained to achieve maximum performance on the transformed images $\mathcal{T}_{S2T}(S2T)$ using labels from $S$. We assess the performance `recovery' of $\mathcal{T}_{S2T}$ by comparing $\mathcal{T}_S(T)$ with $\mathcal{T}_{S2T}(T)$. In practice, data from $T$ would be unlabelled. Our approach ensures that test data from the new domain $T$ is not modified in any way. Additionally, in scenarios where the original model $\mathcal{T}_S$ is deployed, it is likely to have been trained on a large, well-curated, high-quality dataset; we cannot assume similar would be available for each new test domain.
Our model-agnostic unsupervised DA is validated on two problems: i) proof-of-concept showing recovery of a classifier's performance on digit recognition, ii) classification and regression tasks with real-world, multi-site T1-weighted brain MRI.
\begin{align}
    \mathcal{L}_{disT} &= \tfrac{1}{2} \left[ \mathcal{L}_{bce}(D_T(S2T), 0) + \mathcal{L}_{bce}(D_T(T), 1)\right].
    \label{eq:dis_loss_1}\\
    \mathcal{L}_{istn}^{S2T} &= \mathcal{L}_{bce}(D_T(S2T), 1) + \tfrac{1}{2}\lambda \left\|T2T - T\right\|_{1}.
    \label{eq:istn_loss_1}\\
    \mathcal{L}_{istn}^{S2T} &= \mathcal{L}_{bce}(D_S(T2S), 0) + \tfrac{1}{2}\lambda \left\|S2S - S\right\|_{1} + \lambda \left\|S2T2S - S\right\|_{1}\label{eq:istn_cycleloss_a2b}.
\end{align}
\section{Materials}
\subsection{Proof-of-concept: Morpho-MNIST Experiments}
\begin{figure}[t]
	\includegraphics[width=\linewidth]{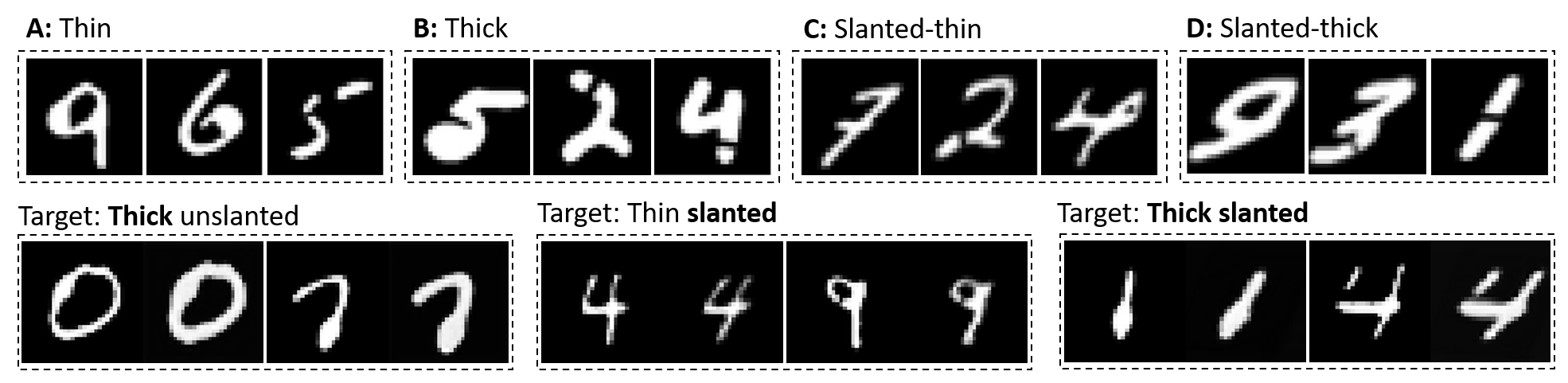}
	\caption{(Top) Examples from Morpho-MNIST datasets from domains (left-to-right) $A$ thin un-slanted digits; $B$ thickened digits; $C$ slanted digits; $D$ thickened \textit{and} slanted digits. Each contains `healthy', `fractured' and `swollen' classes. (Bottom) Examples of source domain images before (left) and after (right) ISTN-transformation showing ISTN recovery of appearance and shape changes.}
		\label{fig:fig2}
\end{figure}

\textbf{Data.} Morpho-MNIST is a framework that enables applying medically-inspired perturbations, such as local swellings and fractures, to the well-known MNIST dataset \cite{castro2019morphomnist}. The framework also allows us to control transformations to obtain thickening and shearing of the original digits. We first create a dataset with three classes: `healthy' digits with no transformations; `fractured' digits with a single thin disruption and `swollen' digits which exhibit a localized, tumor-like abnormal growth. A digit is only either fractured or swollen, not both. We specify a set of `thin' digits (2.5 pixels across) to be source domain $A$. To simulate domain shift, we create three more datasets---domain $B$: thickened, 5.0 pixels digits; domain $C$: slanted digits created by shearing the image by 20--25$^{\circ}$ and domain $D$: thickened-slanted digits at 5.0 pixels and 20--25$^{\circ}$ shearing. Datasets $B$--$D$ contain the same three classes as $A$, while each set has its own data characteristics simulating different types of domain shift. All images are single-channel and $28\times28$ pixels. Figure~\ref{fig:fig2} shows some visual examples.
\\
\\
\textbf{Task.} The downstream task in this experiment is a 3-class classification problem: `healthy' vs.\ `fractured' vs.\ `swollen'. We train a small, fully-convolutional classifier to perform the classification on domain $A$. We use ISTNs to retrain the classifier on transformed images $A2B$, $A2C$, and $A2D$, and evaluate each on their corresponding test domains $B$, $C$, and $D$.

We run training for 100 epochs and perform grid search to find suitable hyper-parameters including learning rate, trade-off $\lambda$ and the control-point spacing of the B-spline STN. We conduct experiments using ITN only, STN only and combinations of affine and B-spline ISTNs to determine the best model for the task. We also consider both transfer directions, switching the roles of source and target domains.

\subsection{Application to Brain MRI Experiments}

We apply the same methodology to a real-world domain shift problem where we observe a significant drop in prediction accuracy when naively training on one site and testing on another without any DA. We utilise 3D brain MRI from two sites that employ similar but not identical imaging protocols.
\\

\noindent\textbf{Data.} We construct two datasets of T1-weighted brain MRI from subjects with no reported pathology, where $n=565$ are taken from the Cambridge Centre for Ageing and Neuroscience study (Cam-CAN) \cite{shafto2014cambridge,taylor2017cambridge} and $n=689$ from the UK Biobank imaging study (UKBB) \cite{sudlow2015,miller2016,alfaro-almagro2018}. From each site, 450 subjects are used for training and the remainder for testing. The UKBB dataset contains equal numbers of male and female subjects between the ages of 48 and 71 ($\mu=59.5$). In the classification task, to simulate the effect of population shift our Cam-CAN dataset has a wider age range (30--87, $\mu=57.9$) but maintains the male-to-female ratio. We match the age range of both datasets in the regression task, limiting DS only to the more subtle scanner effects.
UKBB images were acquired at the UKBB imaging centre, and Cam-CAN images were acquired at the Medical Research Council Cognition and Brain Sciences Unit in Cambridge, UK. Both sites acquire 1\,mm isotropic images using the 3D MPRAGE pulse sequence on Siemens 3\,T scanners with a 32-channel receiver head coil and in-plane acceleration factor 2. Appendix \ref{appendix:appendixA} presents the acquisition parameters that differ between the two sites. We note that generally the acquisition parameters of both sites are similar, and the images cannot be easily distinguished visually.
For pre-processing, all images are affinely aligned to MNI space, skull-stripped, bias-field-corrected, and intensity-normalised to zero mean unit variance within a brain mask. Voxels outside the mask are set to 0. Images are passed through a $\tanh$ function before being consumed by the networks.
\\
\\
\noindent\textbf{Task.} We consider two prediction tasks, namely sex classification and age regression using the UKBB and Cam-CAN sets, each once as source and once as target domain. The task networks are retrained on the transformed images produced by the ISTN and evaluated on the corresponding target domain. 

\section{Experimental Results}
\label{sec:results}
\begin{table}[t]
\vspace{-.5em}
\sffamily
\caption{\fontsize{10pt}{11pt}\selectfont{3-class classification results on MorphoMNIST. Images transferred from classifier domain $A$: `thin unslanted' to three target domains. Accuracies shown for classifiers retrained on the ISTN output from scratch ($\mathrm{Acc_{s}}$) and finetuned ($\mathrm{Acc_{f}}$). $\Delta$ is model improvement from baseline. Control-point spacings indicated for B-Spline STNs. First row is the original classifier without DA.}}
\label{tab:table1}
\scriptsize
    \centering
\begin{minipage}[t]{\linewidth}
    \centering
    \renewcommand{\arraystretch}{1.10}
    \renewcommand{\tabcolsep}{1.5pt}
    \begin{tabular}{c|c|cccc|cccc|cccc}
    \hline
    \multicolumn{2}{c|}{\textbf{Target}} & \multicolumn{4}{c|}{\textbf{Thick} Unslanted} & \multicolumn{4}{c|}{Thin \textbf{Slanted}} & \multicolumn{4}{c}{\textbf{Thick Slanted}}\\
    \hline
    \textbf{ITN} & \textbf{STN} &
    \multicolumn{1}{c}{\textbf{Acc\textsubscript{s}}} & \multicolumn{1}{c}{$\Delta$} &
    \multicolumn{1}{c}{\textbf{Acc\textsubscript{f}}} & \multicolumn{1}{c|}{$\Delta$} &
    \multicolumn{1}{c}{\textbf{Acc\textsubscript{s}}} & \multicolumn{1}{c}{$\Delta$} &
    \multicolumn{1}{c}{\textbf{Acc\textsubscript{f}}} & \multicolumn{1}{c|}{$\Delta$} &
    \multicolumn{1}{c}{\textbf{Acc\textsubscript{s}}} & \multicolumn{1}{c}{$\Delta$} &
    \multicolumn{1}{c}{\textbf{Acc\textsubscript{f}}} & \multicolumn{1}{c}{$\Delta$}\\
    \hline
    no & no & \multicolumn{4}{c|}{41.2} & \multicolumn{4}{c|}{45.7} & \multicolumn{4}{c}{32.8} \\
    \hline
    yes & no & \textbf{79.0} & \textbf{37.8} & \textbf{83.3} & \textbf{42.1} & 83.4 & 37.7 & 83.3 & 37.6 & \textbf{82.4} & \textbf{49.6} & \textbf{84.6} & \textbf{51.8}\\
    \hline
    no & \textbf{Affine} & 52.4 & 11.2 & 68.9 & 27.7 & 92.4 & 46.7 & 93.0 & 47.3 & 54.8 & 22.0 & 64.8 & 32.0 \\
    no & \textbf{B-spline (4)} & 39.0 & -2.2 & 54.4 & 13.2 & 92.1 & 46.4 & 93.1 & 47.4 & 36.0 & 3.2 & 57.2 & 24.4 \\
    no & \textbf{B-spline (8)} & 49.2 & 8.0 & 61.5 & 20.3 & 92.5 & 46.8 & 92.3 & 46.6 & 37.0 & 4.2 & 61.8 & 29.0 \\
    \hline
    yes & \textbf{Affine} & 78.8 & 37.6 & 77.1 & 35.9 & 86.7 & 41.0 & 88.4 & 42.7 & 81.9 & 49.1 & 83.1 & 50.3 \\
    yes & \textbf{B-spline (4)} & 66.3 & 25.1 & 75.8 & 34.6 & \textbf{92.7} & \textbf{47.0} & 91.0 & 45.3 & 79.3 & 46.5 & 82.7 & 49.9 \\
    yes & \textbf{B-spline (8)} & 69.5 & 28.3 & 77.2 & 36.0 & 91.8 & 46.1 & \textbf{93.4} & \textbf{47.7} & 79.0 & 46.2 & 80.8 & 48.0 \\
    \end{tabular}%
    \renewcommand{\arraystretch}{1}
    \renewcommand{\tabcolsep}{6pt}
\end{minipage}
\end{table}%

\textbf{Morpho-MNIST.} Quantitative results for the synthetic experiments are summarized in Table~\ref{tab:table1}. ITNs are able to harmonize local appearance such as thickness between source and target domains, while STNs perform well in recovering shape variations such as slant. Where both thickness and slant are varied between source and target domains, we note an ITN-only performs as well (or slightly better) than a joint ISTN, suggesting that thickness is more important for the classification task. In Fig. \ref{fig:fig2} we show visual results on how the ISTNs are able to recover both appearance and shape differences between domains.

\begin{table}[t]
\sffamily
\caption{\fontsize{10pt}{11pt}\selectfont Sex classification results on 3D Brain MRI}
\label{tab:table2}
\scriptsize
    \centering
\begin{minipage}[t]{\linewidth}
    \centering
    \renewcommand{\arraystretch}{1.10}
    \renewcommand{\tabcolsep}{1.5pt}
    \begin{tabular}{c|c|cccc|cccc|cccc|cccc}
    \hline
    \multicolumn{2}{c|}{\textbf{Source}} & \multicolumn{8}{c|}{\textbf{UKBB}} & \multicolumn{8}{c}{\textbf{Cam-CAN}}\\
    \hline
    \multicolumn{2}{c|}{\textbf{Method}} & \multicolumn{4}{c|}{\textbf{Uni-ISTN}} & \multicolumn{4}{c|}{\textbf{CycleGAN Bi-ISTN}} & \multicolumn{4}{c|}{\textbf{Uni-ISTN}} & \multicolumn{4}{c}{\textbf{CycleGAN Bi-ISTN}}\\
    \hline
    \textbf{ITN} & \textbf{STN} & \multicolumn{1}{c}{\textbf{Acc\textsubscript{s}}} & \multicolumn{1}{c}{\textbf{$\Delta$}} & \multicolumn{1}{c}{\textbf{Acc\textsubscript{f}}} & \multicolumn{1}{c|}{\textbf{$\Delta$}} & \multicolumn{1}{c}{\textbf{Acc\textsubscript{s}}} & \multicolumn{1}{c}{\textbf{$\Delta$}} & \multicolumn{1}{c}{\textbf{Acc\textsubscript{f}}} & \multicolumn{1}{c|}{\textbf{$\Delta$}} & \multicolumn{1}{c}{\textbf{Acc\textsubscript{s}}} & \multicolumn{1}{c}{\textbf{$\Delta$}} & \multicolumn{1}{c}{\textbf{Acc\textsubscript{f}}} & \multicolumn{1}{c|}{\textbf{$\Delta$}} & \multicolumn{1}{c}{\textbf{Acc\textsubscript{s}}} & \multicolumn{1}{c}{\textbf{$\Delta$}} & \multicolumn{1}{c}{\textbf{Acc\textsubscript{f}}} & \multicolumn{1}{c}{\textbf{$\Delta$}}\\
    \hline
    no & no & \multicolumn{4}{c|}{54.8} & \multicolumn{4}{c|}{54.8} & \multicolumn{4}{c|}{64.3} & \multicolumn{4}{c}{64.3}\\
    \hline
    yes & no & 79.1 & 24.3 & 72.2 & 17.4 & \textbf{80.0} & \textbf{25.2} & 80.8 & 26.0 & \textbf{86.2} & 21.9 & 78.2 & 13.9 & 80.8 & 16.5 & 79.9 & 15.6 \\
    \hline
    yes & \textbf{Affine} & \textbf{80.9} & \textbf{26.1} & 75.7 & 20.9 & 70.4 & 15.6 & \textbf{82.4} & \textbf{27.6} & 79.9 & 15.6 & 79.1 & 14.8 & \textbf{82.4} & \textbf{18.1} & 72.0 & 7.7 \\
    yes & \textbf{B-spline (8)} & 78.3 & 23.5 & 76.5 & 21.7 & 79.1 & 24.3 & 78.7 & 23.9 & 80.3 & 16.0 & \textbf{84.5} & \textbf{20.2} & 78.7 & 14.4 & \textbf{80.8} & \textbf{16.5} \\
    yes & \textbf{B-spline (16)} & 80.0 & 25.2 & \textbf{78.3} & \textbf{23.5} & 73.0 & 18.2 & 67.8 & 13.0 & 85.4 & 21.1 & 84.1 & 19.8 & 67.8 & 3.5 & 68.6 & 4.3 \\
    \end{tabular}%
    \renewcommand{\arraystretch}{1}
    \renewcommand{\tabcolsep}{6pt}
\end{minipage}
\end{table}%
\textbf{Brain MRI.} Quantitative results are summarized in Tables \ref{tab:table2} and \ref{tab:table3}. The sex classifier trained and tested on UKBB achieves 84.3\% accuracy. This drops to 54.8\% when tested on Cam-CAN. Similarly, training and testing on Cam-CAN yields 91.6\%, dropping to 64.3\% when testing on UKBB. Using ISTNs for domain adaptation, and retraining the classifiers increases the accuracy substantially on Cam-CAN from 54.8\% to 80.9\%, and on UKBB from 64.3\% to 86.2\%, which is close to the single-site performance. Training the classifier from scratch performs similarly well to fine-tuning. Bidirectional training with CycleGAN seems not to provide substantial improvements over the simpler unidirectional scheme. The ISTNs are able to overcome some of the acquisition and population shifts between the two domains.
\begin{table}[t]
\vspace{-.5em}
\begin{minipage}[t]{0.5\linewidth}
\sffamily
\captionof{table}{\footnotesize Age regression results on 3D Brain MRI. $\mathrm{MAE_{s}}$ is the task model retrained from scratch.}
\setlength{\tabcolsep}{4pt}
\label{tab:table3}
\scriptsize
    \centering
    \renewcommand{\arraystretch}{1.10}
    \renewcommand{\tabcolsep}{1.5pt}
    \begin{tabular}{c|c|cc|cc}
    \hline
    \multicolumn{2}{c|}{\textbf{Source}} & \multicolumn{2}{c|}{\textbf{UKBB}} & \multicolumn{2}{c}{\textbf{Cam-CAN}}\\
    \hline
    \multicolumn{2}{c|}{\textbf{Method}} & \multicolumn{2}{c|}{\textbf{Uni-ISTN}} & \multicolumn{2}{c}{\textbf{Uni-ISTN}}\\
    \hline
    \textbf{ITN} & \textbf{STN} & \multicolumn{1}{c}{\textbf{MAE\textsubscript{s}}} & \multicolumn{1}{c|}{\textbf{$\Delta$}} & \multicolumn{1}{c}{\textbf{MAE\textsubscript{s}}} & \multicolumn{1}{c}{\textbf{$\Delta$}} \\
    \hline
    no & no & \multicolumn{2}{c|}{5.13} & \multicolumn{2}{c}{4.61} \\
    \hline
    yes & no & 4.71 & 0.42 & \textbf{4.57} & \textbf{0.04} \\
    \hline
    yes & \textbf{Affine} & \textbf{4.58} & \textbf{0.55} & 5.00 & -0.39 \\
    yes & \textbf{B-spline (16)} & 5.06 & 0.07 & 4.90 & -0.29 \\
    \end{tabular}%
    \renewcommand{\arraystretch}{1}
    \renewcommand{\tabcolsep}{6pt}
\end{minipage}
\hfill
\begin{minipage}[t]{0.45\linewidth}
\centering
\captionof{figure}{\footnotesize Examples of (left-to-right) source domain, transformed ISTN output and difference image.}
\label{fig:fig3}
\includegraphics[width=\linewidth]{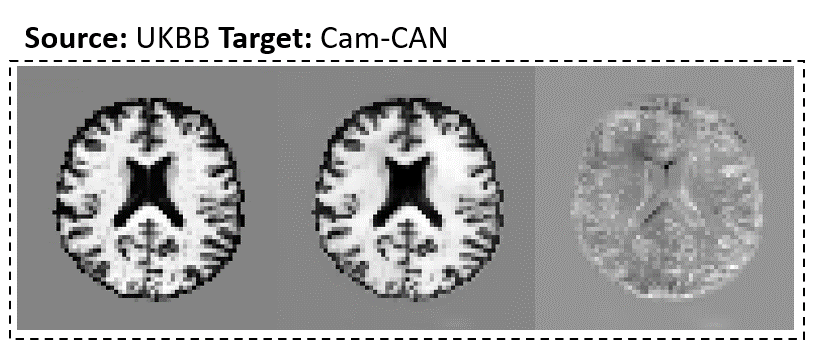}
\end{minipage}
\end{table}
The age regressor trained and tested on UKBB achieves mean absolute error (MAE) of 4.25 years increasing to 5.13 when evaluated on Cam-CAN. The regressor trained and tested on Cam-CAN yields 4.10 years MAE increasing to 4.61 when tested on UKBB. Despite the initially smaller drop in performance for age regression, ISTNs still improve performance. The UKBB-trained regressor recovers to 4.58 years MAE and the Cam-CAN-trained one to 4.56 years. Note, we had limited the population shift here by constraining the age range, thus the recovery is likely due to a reduction in acquisition shift.
\section{Conclusion}
\label{sec:conclusion}

We explored adversarially-trained ISTNs for model-agnostic domain adaptation. The learned image-level transformations help explainability, as the resulting images can be visually inspected and checked for plausibility (cf.\ Fig.~\ref{fig:fig3}). Further interrogation of deformations fields also adds to explainability, \textit{e.g.} Appendix \ref{appendix:appendixB}. Image-level DA seems suitable in cases of subtle domain shift caused by acquisition and population differences in multi-center studies. Predictive performance approached single-site accuracies. The choice of STN and control-point spacings may need to be carefully considered for specific use cases. An extension of our work to many-sites may be possible by simultaneously adapting to multiple sites. A quantitative comparison to feature-level DA would be a natural next step for future work. Another interesting direction could be to integrate the ISTN component in a fully end-to-end task-driven optimisation, where the ISTN and the task network are trained jointly.

\paragraph{\textbf{Acknowledgements:}} RR funded by KCL \& Imperial EPSRC CDT in Medical Imaging (EP/L015226/1) and GlaxoSmithKline; This research received funding from the European Research Council (ERC) under the European Union's Horizon 2020 research and innovation programme (grant agreement No 757173, project MIRA, ERC-2017-STG). DCC is supported by the EPSRC Centre for Doctoral Training in High Performance Embedded and Distributed Systems (HiPEDS, grant ref EP/L016796/1). The research was supported in part by the National Institutes of Health, Clinical Center.

\bibliographystyle{splncs}
\bibliography{cites}

\begin{appendix}
\chapter*{Supplementary Material}

\section{Acquisition Parameters.}
\label{appendix:appendixA}
\begin{table}[h]
\sffamily
    \caption{\fontsize{10pt}{10pt}\selectfont{Acquisition parameters for the multi-site brain MRI datasets.} \label{tab:appendixA}}
    \centering
    \small
    \begin{tabular}[!tb]{l@{\hskip 6pt}lccccc}
        \hline
        \textbf{Site}  & \textbf{Scanner} & \textbf{TR (ms)} & \textbf{TE (ms)} & \textbf{TI (ms)} & \textbf{TA (s)} & \textbf{FOV (mm)} \\ \hline
        \textbf{Cam-CAN} & Siemens TIM Trio & 2250  & 2.99  & 900   & 272   & 256x240x192 \\
        \textbf{UKBB}    & Siemens Skyra    & 2000  & 2.01   & 880   & 294   & 208x256x256
    \end{tabular}
\end{table}

\clearpage
\section{ISTN Transformation Visualization.}
\label{appendix:appendixB}
\begin{figure}[h!]
	\includegraphics[width=\linewidth]{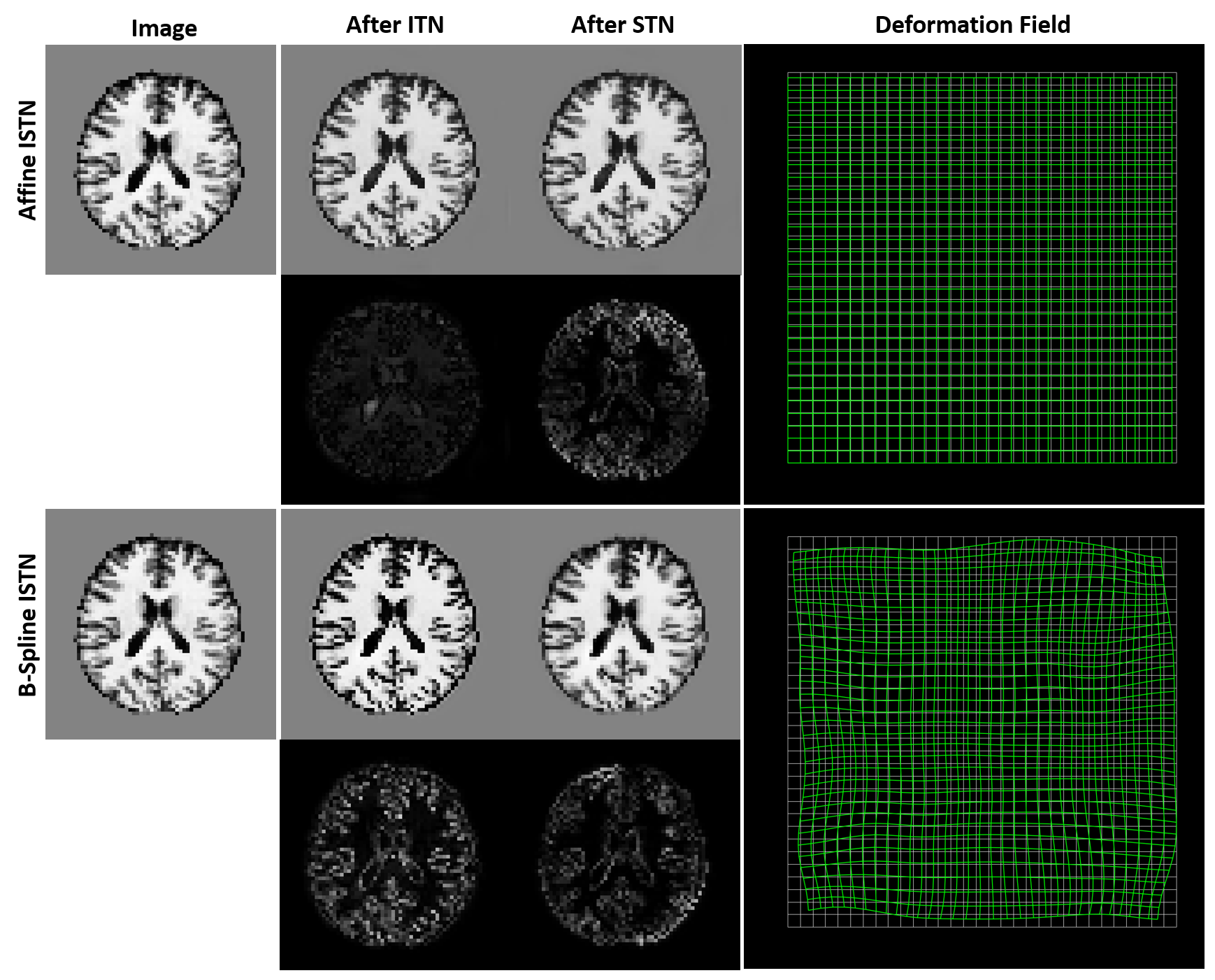}
	\caption{The original image (left) passes through the ISTN. The transformations applied by the ITN and subsequently by the STN are visualized by showing difference images. The transformation applied by the STN can also be visualized as a spatial deformation field (right). This is shown for the Affine (top) and B-Spline (bottom) STNs.}
		\label{fig:appendixB}
\end{figure}

\clearpage
\section{Morpho-MNIST Architectures.}
\label{appendix:appendixC}
\begin{table}[!h]
\sffamily
\centering
\caption{\footnotesize ITN (left) and discriminator (right) architectures for Morpho-MNIST experiments. $nf$: number of channels, $k$: square kernel size, $s$: stride, $in$ and $out$: layer input and output dimensions, $N$: normalization (BN: batch normalization, IN: instance normalization), $D$: Dropout keep-rate, $A$: activation function. `up' is composed of bilinear upsampling followed by zero-padding of 1 and convolution shown in the table.}
\vspace{0.2cm}
\label{tab:appendixC}
\scriptsize
    \renewcommand{\arraystretch}{1.10}
    \renewcommand{\tabcolsep}{3pt}
        \begin{tabular}[t]{rrrrrllccl}
        \multicolumn{10}{c}{\textbf{ITN Architecture - Morpho-MNIST}} \\
        \midrule 
        \textbf{layer} & \textbf{nf} & \textbf{k} & \textbf{s} & \textbf{p} & \textbf{in} & \textbf{out} & \textbf{N} & \multicolumn{1}{l}{\textbf{D}} & \textbf{A} \\
        \midrule 
        \textbf{in}     & - & - & - & - & {[}1,28,28{]} & -                 & - &     & -    \\
        \hline
        \textbf{conv}   & 16 & 3 & 1 & 1 & {[}1,28,28{]} & {[}16,28,28{]} & BN & - & ReLU \\
        \hline
        \textbf{conv}   & 32& 3 & 2 & 1 & {[}16,28,28{]} & {[}32,14,14{]}& BN& - & ReLU \\
        \textbf{conv}   & 64& 3 & 2 & 1 & {[}32,14,14{]}& {[}64,7,7{]}  & BN& - & ReLU \\
        \hline
        \textbf{conv}   & 128& 3 & 1 & 1 & {[}64,7,7{]}  & {[}128,7,7{]}  & BN & - & ReLU \\ 
        \textbf{conv}   & 64& 3 & 1 & 1 & {[}128,7,7{]}  & {[}64,7,7{]}  & BN & - & ReLU \\
        \hline
        \textbf{up}     & 32& 3 & 1 & 1 & {[}64,7,7{]}  & {[}32,14,14{]}& BN& - & ReLU \\
        \textbf{up}     & 16 & 3 & 1 & 1 & {[}32,14,14{]}& {[}16,28,28{]} & BN& - & ReLU \\
        \hline
        \textbf{up}     & 1 & 3 & 1 & 1 & {[}16,28,28{]}& {[}1,28,28{]} & -& - & tanh \\
        \hline
        \textbf{out}    & -  & - & - &   & -            & {[}1,28,28{]} & -  & -   & - \\
    \midrule
        \multicolumn{10}{c}{\textbf{Discriminator Architecture - Morpho-MNIST}}                                                                    \\ \hline
        \textbf{layer} & \textbf{nf} & \textbf{k} & \textbf{s} & \textbf{p} & \textbf{in} & \textbf{out} & \textbf{N} & \multicolumn{1}{l}{\textbf{D}} & \textbf{A} \\ \hline
        \textbf{in}     & - & - & - & - & {[}1,28,28{]} & -                 & - &     & -    \\
        \hline
        \textbf{conv}   & 32 & 3 & 1 & 1 & {[}1,28,28{]} & {[}32,28,28{]} & - & - & ReLU \\
        \hline
        \textbf{conv}   & 64& 3 & 2 & 1 & {[}32,28,28{]} & {[}64,14,14{]}& IN& - & ReLU \\
        \textbf{conv}   & 128& 3 & 2 & 1 & {[}64,14,14{]}& {[}128,7,7{]}  & IN& - & ReLU \\
        \textbf{conv}   & 256& 3 & 2 & 1 & {[}128,7,7{]}& {[}256,4,4{]}  & IN& 0.5 & ReLU \\
        \hline
        \textbf{conv}   & 1& 3 & 2 & 1 & {[}256,4,4{]}& {[}1,1,1{]}  & -& - & sigmoid \\
        \hline
        \textbf{out}    & -  & - & - &   & -            & {[}1,1,1{]} & -  & -   & -   \\
    \midrule
        \multicolumn{10}{c}{\textbf{3-Class Classifier Architecture - Morpho-MNIST}}                                                                    \\ \hline
        \textbf{layer} & \textbf{nf} & \textbf{k} & \textbf{s} & \textbf{p} & \textbf{in} & \textbf{out} & \textbf{N} & \multicolumn{1}{l}{\textbf{D}} & \textbf{A} \\ \hline
        \textbf{in}     & - & - & - & - & {[}1,64,64,64{]} & -                 & - &     & -    \\
        \hline
        \textbf{conv}   & 16 & 3& 1 & 1 & {[}1,24,24{]} & {[}16,24,24{]} & - & - & ReLU \\
        \hline
        \textbf{conv}   & 32& 3 & 2 & 1 & {[}16,14,14{]} & {[}32,7,7{]}& BN& - & ReLU \\
        \textbf{conv}   & 64& 3 & 2 & 1 & {[}32,7,7{]}& {[}64,4,4{]}  & BN& - & ReLU \\
        \textbf{conv}   & 128& 3 & 2 & 1 & {[}64,4,4{]}& {[}128,1,1{]}  & BN& 0.5 & ReLU \\
        \hline
        \textbf{conv}   & 3& 3 & 2 & 0 & {[}128,1,1{]}& {[}3,1,1{]}  & -& - & sigmoid \\
        \hline
        \textbf{out}    & -  & - & - &   & -            & {[}3,1,1{]} & -  & -   & -   
        \end{tabular}
    \renewcommand{\arraystretch}{1.0}
    \renewcommand{\tabcolsep}{6pt}
\end{table}

\clearpage
\section{Brain MRI Architectures.}
\label{appendix:appendixD}
\begin{table}[!h]
\sffamily
\caption{\footnotesize Architectures for Brain MRI experiments. $nf$: number of channels, $k$: square kernel size, $s$: stride, $in$ and $out$: layer input and output dimensions, $N$: normalization (BN: batch or IN: instance normalization), $D$: Dropout keep-rate, $A$: activation function. `up' is composed of linear upsampling, zero-padding and convolution.}
\label{tab:appendixD}
\centering
\scriptsize
    \renewcommand{\arraystretch}{1.10}
    \renewcommand{\tabcolsep}{3pt}
        \begin{tabular}[t]{rrrrrllccl}
        \multicolumn{10}{c}{\textbf{ITN Architecture - Brain MRI}}                                                                    \\ \hline
        \textbf{layer} & \textbf{nf} & \textbf{k} & \textbf{s} & \textbf{p} & \textbf{in} & \textbf{out} & \textbf{N} & \multicolumn{1}{l}{\textbf{D}} & \textbf{A} \\ \hline
        \textbf{in}     & - & - & - & - & {[}1,64,64,64{]} & -                 & - &     & -    \\
        \hline
        \textbf{conv}   & 8 & 3 & 1 & 1 & {[}1,64,64,64{]} & {[}8,64,64,64{]} & BN & - & ReLU \\
        \hline
        \textbf{conv}   & 16& 3 & 2 & 1 & {[}8,64,64,64{]} & {[}16,32,32,32{]}& BN& - & ReLU \\
        \textbf{conv}   & 32& 3 & 2 & 1 & {[}16,32,32,32{]}& {[}32,16,16,16{]}  & BN& - & ReLU \\
        \textbf{conv}   & 64& 3 & 2 & 1 & {[}32,16,16,16{]}& {[}64,8,8,8{]}  & BN& - & ReLU \\
        \hline
        \textbf{conv}   & 64& 3 & 1 & 1 & {[}64,8,8,8{]}  & {[}64,8,8,8{]}  & BN & - & ReLU \\ 
        \textbf{conv}   & 64& 3 & 1 & 1 & {[}64,8,8,8{]}  & {[}64,8,8,8{]}  & BN & - & ReLU \\
        \hline
        \textbf{up}     & 32& 3 & 1 & 1 & {[}64,8,8,8{]}  & {[}32,16,16,16{]}& BN& 0.5 & ReLU \\
        \textbf{up}     & 16& 3 & 1 & 1 & {[}32,16,16,16{]}  & {[}16,32,32,32{]}& BN& 0.5 & ReLU \\
        \textbf{up}     & 8 & 3 & 1 & 1 & {[}16,32,32,32{]}& {[}8,64,64,64{]} & BN& 0.5 & ReLU \\
        \hline
        \textbf{up}     & 1 & 3 & 1 & 1 & {[}8,64,64,64{]}& {[}1,64,64,64{]} & -& - & tanh \\
        \hline
        \textbf{out}    & -  & - & - &   & -            & {[}1,64,64,64{]} & -  & -   & -   \\
    \midrule
        \multicolumn{10}{c}{\textbf{Discriminator Architecture - Brain MRI}}                                                                    \\ \hline
        \textbf{layer} & \textbf{nf} & \textbf{k} & \textbf{s} & \textbf{p} & \textbf{in} & \textbf{out} & \textbf{N} & \multicolumn{1}{l}{\textbf{D}} & \textbf{A} \\ \hline
        \textbf{in}     & - & - & - & - & {[}1,64,64,64{]} & -                 & - &     & -    \\
        \hline
        \textbf{conv}   & 32 & 3 & 1 & 1 & {[}1,64,64,64{]} & {[}32,64,64,64{]} & - & - & ReLU \\
        \hline
        \textbf{conv}   & 64& 3 & 2 & 1 & {[}32,64,64,64{]} & {[}64,32,32,32{]}& IN& - & ReLU \\
        \textbf{conv}   & 128& 3 & 2 & 1 & {[}64,32,32,32{]}& {[}128,16,16,16{]}  & IN& - & ReLU \\
        \textbf{conv}   & 256& 3 & 2 & 1 & {[}128,16,16{]}& {[}256,8,8{]}  & IN& - & ReLU \\
        \textbf{conv}   & 256& 3 & 2 & 1 & {[}256,8,8{]}& {[}256,4,4{]}  & IN& 0.5 & ReLU \\
        \hline
        \textbf{conv}   & 1& 3 & 2 & 1 & {[}256,4,4{]}& {[}1,1,1{]}  & -& - & sigmoid \\
        \hline
        \textbf{out}    & -  & - & - &   & -            & {[}1,1,1{]} & -  & -   & -   \\
    \midrule
        \multicolumn{10}{c}{\textbf{Sex Classifier Architecture - Brain MRI}}                                                                    \\ \hline
        \textbf{layer} & \textbf{nf} & \textbf{k} & \textbf{s} & \textbf{p} & \textbf{in} & \textbf{out} & \textbf{N} & \multicolumn{1}{l}{\textbf{D}} & \textbf{A} \\ \hline
        \textbf{in}     & - & - & - & - & {[}1,64,64,64{]} & -                 & - &     & -    \\
        \hline
        \textbf{conv}   & 8 & 5 & 2 & 2 & {[}1,64,64,64{]} & {[}8,64,64,64{]} & - & - & ReLU \\
        \hline
        \textbf{conv}   & 16& 5 & 2 & 2 & {[}8,64,64,64{]} & {[}16,32,32,32{]}& BN& - & ReLU \\
        \textbf{conv}   & 32& 5 & 2 & 2 & {[}16,32,32,32{]}& {[}32,16,16,16{]}  & BN& - & ReLU \\
        \textbf{conv}   & 64& 5 & 2 & 2 & {[}32,16,16{]}& {[}64,8,8{]}  & BN& 0.5 & ReLU \\
        \textbf{conv}   & 128& 2 & 2 & 2 & {[}64,8,8{]}& {[}128,4,4{]}  & BN& 0.5 & ReLU \\
        \textbf{conv}   & 128& 2 & 2 & 2 & {[}128,4,4{]}& {[}128,1,1{]}  & BN& 0.5 & ReLU \\
        \hline
        \textbf{conv}   & 1& 5 & 1 & 2 & {[}128,1,1{]}& {[}1,1,1{]}  & -& - & sigmoid \\
        \hline
        \textbf{out}    & -  & - & - &   & -            & {[}1,1,1{]} & -  & -   & -   \\
    \midrule
        \multicolumn{10}{c}{\textbf{Age Regressor Architecture - Brain MRI}}                                                                    \\ \hline
        \textbf{layer} & \textbf{nf} & \textbf{k} & \textbf{s} & \textbf{p} & \textbf{in} & \textbf{out} & \textbf{N} & \multicolumn{1}{l}{\textbf{D}} & \textbf{A} \\ \hline
        \textbf{in}     & - & - & - & - & {[}1,64,64,64{]} & -                 & - &     & -    \\
        \hline
        \textbf{conv}   & 16 & 3 & 1 & 1 & {[}1,64,64,64{]} & {[}8,64,64,64{]} & - & - & ReLU \\
        \textbf{MaxPool} & - &  - & 2 & 1 & {[}8,64,64,64{]} & {[}8,32,32,32{]} & - & - & - \\
        \textbf{conv}   & 32& 3 & 2 & 1 & {[}8,32,32,32{]} & {[}32,32,32,32{]}& - & - & ReLU \\
        \textbf{MaxPool} & - &  - & 2 & 1 & {[}32,32,32,32{]} & {[}32,16,16,16{]} & - & - & - \\
        \hline
        \textbf{Linear}   & 128& - & - & - & {[}32*32*32*32,1{]}& {[}128{]}  & - & - & ReLU \\
        \textbf{Linear}   & 64& - & - & - & {[}128{]}& {[}64{]}  & - & - & ReLU \\
        \textbf{Linear}   & 32& - & - & - & {[}64{]}& {[}32{]}  & - & - & ReLU \\
        \textbf{Linear}   & 1& - & - & - & {[}32{]}& {[}1{]}  & - & - &  \\
        \hline
        \textbf{out}    & -  & - & - &   & -            & {[}1{]} & -  & -   & -   
        \end{tabular}
    \renewcommand{\arraystretch}{1}
    \renewcommand{\tabcolsep}{6pt}
\end{table}

\end{appendix}

\end{document}